\documentclass{ws-procs9x6}
\input epsf
\begin{document}

\title{Gravitational Wave Detectors: A report from LIGO-land
\footnote{\uppercase{T}his work is supported by the \uppercase{N}ational \uppercase{S}cience \uppercase{F}oundation, with grant \uppercase{PHY}-0135389.}}

\author{GABRIELA GONZ{\'A}LEZ \\ for the LIGO Science Collaboration}

\address{Department of Physics and Astronomy \\
Louisiana State University\\
202 Nicholson Hall, \\ 
Baton Rouge, LA 70803, USA\\ 
E-mail: gonzalez@lsu.edu}

\maketitle

\abstracts{At the time of this conference, in June 2002, The LIGO
Science Collaboration was getting ready to perform its first Science
Run, where data will be taken with all three LIGO detectors. We
describe here the status of the LIGO detectors as of February 2003,
their performance during the ``Engineering Run'' E7 (Dec 28'01-Jan
14'02) and subsequent Science Runs in 2002/3. We also describe ongoing
efforts on data analysis for setting upper limits of different
gravitational wave sources. }

\section{Introduction}

The LIGO project has two Observatories, in Hanford, Washington and
Livingston, Louisiana; each housing km-long interferometric
gravitational wave detectors. There are twin detectors, 4km long each,
in Livingston and Hanford; in Hanford the project also has a 2km long
detector living in the same vacuum system. 

The LIGO Science Collaboration consists now of 33 institutions, from
all five continents. There are about 450 collaborators, working on
commissioning the present LIGO detectors, preparing for data analysis
and testing it on preliminary data, and doing research and development
on techniques for improving the detectors in the near future.
 
As of June 2002, all three LIGO detectors were working in their final
configurations, and they were about to start the first Science data
taking run. The sensitivity of the detectors was still far from the
one ultimately expected, limited by fundamental noise
sources. However, the preliminary data provides an opportunity for
careful diagnostics of noise sources, and for greasing the wheels of
the data analysis machinery that will be used. Some of these
techniques were used to analyze the data from the ``E7'' Engineering
Run, from December 28, 2001 to January 14, 2002. 

We will describe in this article the basics of the signal detection
scheme used in the LIGO detectors, and present preliminary results
from the analysis on E7 data.

\section{Interferometric gravitational wave detectors}

There are now several interferometric gravitational wave detectors
expected to begin taking data in the next couple of years: GEO600
\cite{GEO600} is a German-British collaboration; VIRGO \cite{VIRGO}, a
French-Italian collaboration; LIGO\cite{LIGO}, an US project where the
science is managed by an international collaboration; and
TAMA300\cite{TAMA}, a Japanese project; ACIGA, an ACIGA, an Australian
project.  TAMA300 has already been taking data. All of these detectors
differ in dimensions, expected sensitivities, and configurations, but
they all share the basic idea of using laser optical interferometry,
Pond-Drever detection scheme and suspended mirrors to detect the small
effects produced by gravitational waves on Earth.

The configuration of the LIGO detectors is that of a power recycled
Michelson interferometer with Fabry Perot arms, as described in
\cite{LIGO}. The arms of the Michelson interferometer are 4km long,
and the optical cavities in each arm have a finesse of about 100. The
power recycling cavity has a finesse of 30. The laser used is a 6W
NdYag Lightwave NPRO, with a wavelength of 1.064 $\mu$m. The frequency
of the light is first stabilized outside the vacuum system with respect to
the length of a reference cavity, and then using a solid pre-mode cleaner. In
vacuum, the gaussian beam is further stabilized in spatial mode and
frequency stability by a triangular suspended mode cleaner.

A power-recycled, Fabry Perot interferometer has four degrees of
freedom that need to be controlled so that the optical cavities are
resonant. Two of these are the length of the resonant arms, and the
light sensed at the antisymmetric and reflected port is used to
generate control signals for the difference and sum of the arm
lengths. The signal at the antisymmetric, or ``dark'' port, which is
proportional to the difference of the arm lengths, is the one most
sensitive to gravitational waves. 

The other two degrees of freedom to be controlled are the
recombination at the beamsplitter to keep the antisymmetric port dark,
and the length of the recycling cavity to keep the optical cavity
resonant. The signals for these degrees of freedom are sensed in a
secondary reflection picked off from the beamsplitter mirror, or one
of the input test masses in the arm cavities.

In all cases, the signals are first generated using a 25 MHz phase
modulation frequency in th e-input laser light, and then demodulating
the photocurrents in each of the three sensed beams, using the
``Pond-Drever'' sensing scheme. The linearity of the signals with
respect to the desired degrees of freedom require that the signals are
held very close to zero, or oscillating by a small fraction of the
wavelength, divided by the cavity finesse, $\sim \lambda/100$. When
this is achieved, it is said f that the interferometer is ``locked''. 

For the arm length difference, we require the test masses to be held
in position to $10^{-13}$ meter within the resonance of the optical
cavities. This level of control has now been achieved in all three
LIGO detectors. The LIGO detectors are expected to be sensitive to
fluctuations in the difference of arm lengths of about $10^{-19}$ m,
if happening at frequencies between 50Hz and 4 kHz.

When the suspended mirrors are not being controlled, they are swinging
through more than one wavelength, so the signals at the photodiodes
are very non-linear. An end-to-end model of the electromagnetic fields
was used to interpret the signals at the photodiodes when the
interferometer is not locked. Using the results of this modeling, a
digital program calculates ``on the fly'' (at a rate of 16 kHz) the
appropriate forces to send to the mirrors and lock the interferometer.

\begin{figure}[ht]
\centerline{\epsfxsize=5in\epsfbox{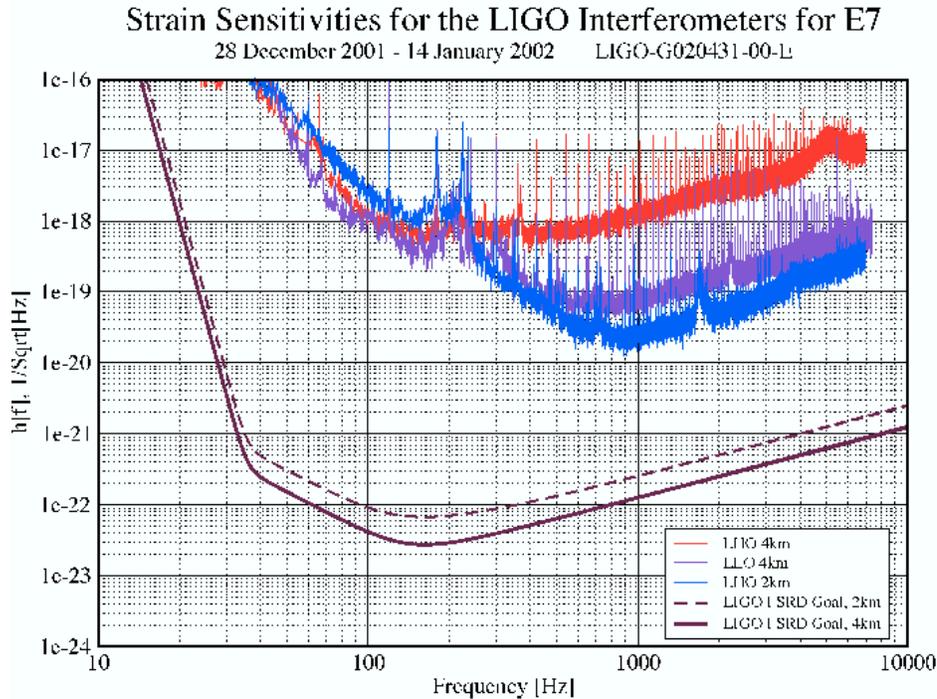}}   
\caption{Sensitivity of LIGO detectors during E7 data taking run. The
solid and dotted lines are the expected sensitivities of the detectors
a the end of the commissioning effort.\label{LIGOE7}}
\end{figure}

As of June 2002, all three LIGO interferometers can lock in the full
recycled mode. However, anthropogenic and natural sources of seismic
noise have a big impact on duty cycle. The mirrors are pushed around
by coils acting on magnets glued to the mirrors' back surface; these
systems have a small dynamic range and sometimes, they cannot apply
the necessary forces to move the mirrors far enough to cancel the
motion of the ground. There are massive, passive seismic isolation
systems installed in the vacuum system to reduce the seismic noise in
the gravitational wave band, however these systems have resonances and
in fact amplify the ground motion at some frequencies. The ground
motion at low frequencies is largest in the Livingston Observatory,
where logging happens nearby. Several active systems are now being
used to improve this situation, and a more sophisticated version
close to what was planned for Advanced LIGO will be installed in 2003.

\section{Data Taking Runs}

Several ``Engineering Runs'' were done to assess the performance of
the LIGO detectors and test the software tools used for searching
gravitational waves. Four groups were created within the LIGO Science
Collaboration, each one looking at methods for searching gravitational
waves of different kinds: Stochastic Background, Continuous Periodic
Waves, Burst Sources and Inspiral Signals. The last Engineering Run
before this conference, ``E7'', took place between Dec 28, 2001 and
January 14, 2002.  The three LIGO detectors were in operation, and
data was also taken in coincidence with the GEO600 detector in
Hannover, Germany, and the ALLEGRO bar detector in Louisiana State
University, USA (50 km away from the LIGO Livingston
detector). Although the data quality for the interferometric detectors
was far from the expected sensitivities when the undergoing
commissioning is finished, a lot was learned from this data taking
run. The duty cycles measured are detailed in Table \ref{E7Table}. The
sensitivities of the LIGO detectors are shown in Fig.\ref{LIGOE7}. 

The LIGO H1 interferometer (4km long in Hanford, Washington), had
achieved a stable lock only a few days before the data taking run
started, so even though its sensitivity was lagging behind the other
LIGO detectors (L1, 4 km long at Livingston, and H2, 2km long at
Hanford), it was the first time that all three LIGO detectors were
running in coincidence. The 2km Hanford detector, H2, started to
suffer problems to stay locked during the latter part of the run, the
duty cycle at the end of the run was much smaller than in the
beginning. The Livingston detector, L1, was operated in an
intermediate configuration, where the arm cavities were
``recombined'', but power recycling was not used. This configuration
is in principle limited by a worse shot noise limit at high
frequencies, since it has less circulating power, but other noise
sources were still dominating the noise budget. Some of the known
noise sources were spurious forces introduced by angular motion
controllers; thermal noise at sensing photodiodes; frequency noise in
the input light to the interferometer; acoustic vibrations of input
periscopes; etc. All the noise sources dominating the LIGO spectra at
the time were identified and suppressed; the sensitivity of all three
detectors at the time of writing this article (February 2003), is only
a factor of a few to ten times above the design sensitivity, and ten
to a hundred times better than the noise figures plotted in
Fig\ref{LIGOE7}.

\begin{table}[ph]
\tbl{Hours of useful locked time and duty cycles of detectors 
participating in the E7 data taking run.\vspace*{1pt}}
{\footnotesize
\begin{tabular}{|c|c|c|c|}
\hline
{}& {} &{} &{} \\[-1.5ex]
{}& LIGO detectors & LIGO and GEO & L1 and ALLEGRO\\[1ex]
\hline
Recorded time:& 408 hrs & 334 hrs & 222 hrs \\[1ex]
\hline
{}& {} &{} &{}\\[-1.5ex]
Single&
     L1: 249 hrs (62\%) & L1: 197 hrs (59\%) & L1: 91 hrs (41\%) \\[1ex]
detectors & H1: 231 hrs (57\%) & H1: 109 hrs (63\%) & ALLEGRO: 165 hrs (74\%) \\[1ex] 
{}& H2: 157 hrs (39\%) & H2: 109 hrs (32\%) & \\ [1ex]
  &                 & GEO: 211 hrs (63\%) & \\[1ex]
\hline
Coincidence & 72 hrs (18\%) & 26 hrs (8\%) & 78 hrs (35\%) \\[1ex]
\hline
\end{tabular}\label{E7Table} }
\vspace*{-13pt}
\end{table}

A figure of merit derived from the noise spectra is the maximum
distance from which a signal from an optimally oriented binary neutron
star system could be detected with a signal-to-noise ratio of 8. This
number varies with the quality of the mirror alignment in the
interferometers, but it was in average 0.8 kpc in H1, 3.5 kpc in L1
and 11 kpc in H2.

A significant amount of attention was devoted to looking into
auxiliary channels to find correlations with apparent signals in the
gravitational wave channels. In L1, it was found that a channel
indicating laser amplitude fluctuations showed a a very important
correlation with the gw signal; it was consequently used as a ``veto''
channel. In H2, the detector with best sensitivity, it was found that
many time s after locking, and sometimes during locked times,
vibrations of the ``violin'' modes of the suspension wires were
excited by the control system. A tool was developed to calculate the
power in a small frequency band containing the frequencies of the
violin modes, and when this power exceeded certain threshold, the data
was considered to be vetoed. The exercise of studying the signals in
detail, and finding the origin of the obvious instrumental artifacts,
was one of the most useful lessons learned from the E7 data analysis.

The GEO600 detector was also operated in coincidence with the LIGO
detectors during E7. It was also operating with a different
configuration than designed: it used power recycling but not signal
recycling. The GEO detector had the best duty cycle of all
interferometers, it improved to an amazing 98-99\% by the final days
of the run. 

\begin{figure}[ht]
\centerline{\epsfxsize=4in\epsfbox{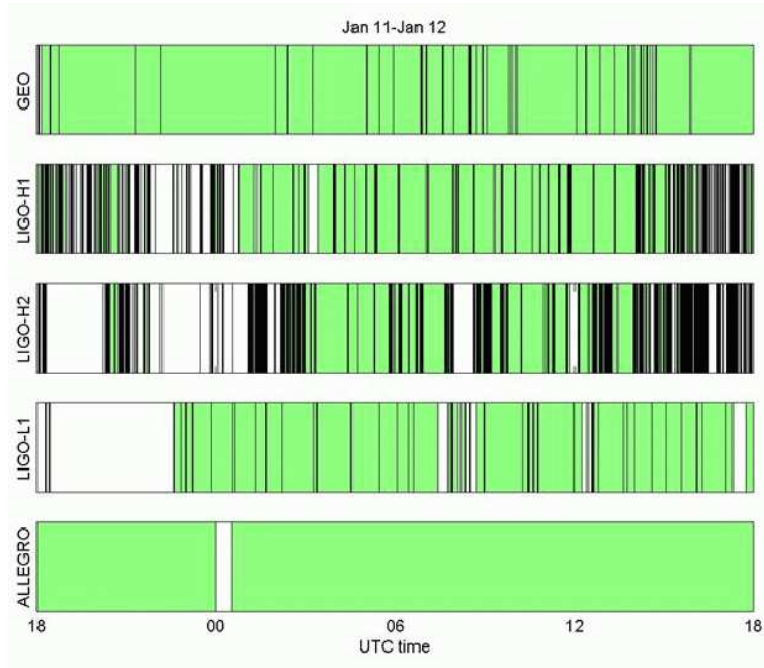}}   
\caption{Status of detectors during 24 hours of E7 (Jan 11 18:00
UTC-Jan 12 18:00 UTC) Green means ON, white is OFF, with the regions
having black boundaries.\label{Jan11}}
\end{figure}

The ALLEGRO resonant detector operated during the last week of E7,
from January 7 to January 14. It made use for the first time of its
ability to rotate orientations; it started operating aligned with the
International Gravitational Event Collaboration (IGEC)\cite{IGEC},
then rotated to align the bar with the X-arm of the LIGO Livingston
detector on January 11; and then on January 12 to a direction at 45
degrees with both X and Y arms of the LIGO LLO detector. This strategy
allows for a comparison of the signals correlations in an ON/OFF state
for searches of stochastic background \cite{FinnLazz} .

A plot of the ON/OFF status of all five detectors participating in E7
is shown in Figure \ref{Jan11}.

\section{\it Note added in Proof}

At the time of writing this article, data from the first science run
(August 23-September 9, 2002) is being analyzed. The S1 run was also
in coincidence with GEO, although not with ALLEGRO. The data showed
considerably improvement with respect to E7: the LIGO detectors were
now sensitive to binary inspiral neutron star systems in the galaxy,
at distances between 50 kpc (H2) and 150 kpc (L1). This range allows a
fine survey of the galactic source with coincident detectors.
There were 96 hours of triple coincidence time, and 131 hours of
double coincidence between L1 and H1. The duty cycle was again limited
by daytime seismic noise limiting the LIGO Livingston detector locking
ability. The analysis done by the four upper limit groups shows very
promising results, including the efficient use of two detectors on
line for stochastic background, and burst and inspiral sources.

In February 2003, a second Science run, to be 60 days long, has
started, again with improved sensitivities with respect to S1: the
LIGO Hanford detectors are bot now sensitive to inspiraling binary
neutron stars at 600 kpc (H1), 400 kpc (H2) and 1.5 Mpc (L1). With
these ranges, the Large Magellanic Cloud can be ``seen'' with triple
coincidence, and for the first time, an interferometric detector will
be sensitive to sources in the Andromeda Galaxy.

\section{Conclusions}

The LIGO interferometers are all operating, with a sensitivity getting
closer to the designed one at a fast pace. By the time of the NEB-X
conference, an important threshold in operations had been crossed,
with data being taken in all three LIGO detectors, in coincidence with
GEO and ALLEGRO; and another threshold was being crossed, glancing at
Andromeda with a gravitational wave detector for the first time. The
rapid progress of the instruments' performance is promising a good
start of a new era of gravitational wave astronomy.

\end{document}